\documentclass[10pt,conference]{IEEEtran}

\usepackage{graphicx}
\usepackage[utf8]{inputenc}

\usepackage{multirow}
\usepackage{hyperref}
\usepackage[lofdepth,lotdepth]{subfig}
\usepackage{url}
\usepackage{multirow}
\usepackage{multicol}

\usepackage{supertabular,booktabs} 

\usepackage[flushmargin]{footmisc} 
\usepackage{enumitem} 
\usepackage{cite}

\IEEEoverridecommandlockouts
\newcommand*{\nolink}[1]{%
  \begin{NoHyper}#1\end{NoHyper}%
}

\title{GDPR-inspired IoT Ontology enabling Semantic Interoperability, Federation of Deployments and Privacy-Preserving Applications}

\author{
    \IEEEauthorblockN{Rachit Agarwal\IEEEauthorrefmark{1}\IEEEauthorrefmark{4},
        Tarek Elsaleh\IEEEauthorrefmark{2}, 
        Elias Tragos\IEEEauthorrefmark{3} 
    }
    \IEEEauthorblockA{
        \IEEEauthorrefmark{1}Inria-Paris, France, 
        \IEEEauthorrefmark{4}CSE, IIT Kanpur, India,
        \IEEEauthorrefmark{2}University of Surrey, UK, 
        \IEEEauthorrefmark{3}Insight Center for Data Analytics-UCD, Ireland\\
        Email: \IEEEauthorrefmark{1}{rachit.agarwal@inria.fr},
                \IEEEauthorrefmark{2}{t.elsaleh@surrey.ac.uk}, \IEEEauthorrefmark{3}{elias.tragos@insight-centre.org}
    }
}

\begin{document}

\maketitle

\begin{abstract} 
Testing and experimentation are crucial for promoting innovation and building systems that can evolve to meet high levels of service quality. IoT data that belong to users and from which their personal information can be inferred are frequently shared in the background of IoT systems with third parties for experimentation and building quality services. This data sharing raises privacy concerns especially since in most cases the data are gathered and shared without the user's knowledge or explicit consent or for different purposes than the one for which the data were initially gathered. With the introduction of GDPR, IoT systems and experimentation platforms that federate data from different deployments, testbeds and data providers must be privacy-preserving. The wide adoption of IoT applications in scenarios ranging from smart cities to Industry 4.0 has raised concerns with respect to the privacy of users' data collected using IoT devices. Many experimental smart city applications are also using crowdsourcing data. Inspired by the GDPR requirements, we propose an IoT ontology built using available standards that enhances privacy, enables semantic interoperability between IoT deployments and supports the development of privacy-preserving experimental IoT applications.  On top, we propose recommendations on how to efficiently use the ontology within IoT testbed and federating platforms. Our ontology is validated for different quality assessment criteria using standard validation tools. We focus on ``experimentation" without loss of generality, because it covers scenarios from both research and industry, that are directly linked with innovation. 

\end{abstract}


\vspace{-0.3cm}
\section{Introduction}

The role of experimentation platforms for IoT data and service providers is crucial for building systems that can evolve to meet high levels of service quality and {promote innovation}. These systems should be highly resilient and include feedback mechanisms that provide a quality assessment of different aspects of the data ranging from annotation compliance to the consistency of frequency. To have an end-to-end testing environment that covers the whole data flow from source to consumer, employing testing or staging instances is essential not only for internal testing within the testbed organisation but also for offering external testing services to third parties. 

IoT experimentation assumes that a testbed/data provider provides the collected data to third parties (i.e., researchers) for running experiments. This requires sharing the data with external parties that are not the data owners. Thereby raising privacy concerns if external parties mishandle the data. The latest European Union (EU) directive for data protection (GDPR - General Data Protection Regulation~\cite{gdpr}) sets strict guidelines on the collection and processing of user data. IoT systems are required to be compliant with these guidelines because, in most cases, they handle user-generated data. 

For example, consider a user participating in a crowd-sourcing based environmental monitoring testbed by having installed an application on his smartphone~\cite{Valerie2016}. A third party having access to the data produced by the smartphone and associated with the application may be able to interpret private user information such as the user's location at any given time, or his home/work location. Thus, one important aspect of the provided data is its privacy sensitivity and the context in which the IoT devices (such as smartphones) capture it. 

For a multi-domain experimentation facility, it is important to provide a flexible means to configure the level of data privacy, depending on the association with real-world entities, and the purpose of data processing. With the introduction of GDPR the need to have a data model that reflects the concepts defined by these laws is essential. This requires modelling the roles of different actors, the consent mechanism, and the data owner's permissions on what third parties can consume.

Most of the past activities in IoT experimentation were only focused on developing the fundamental technologies to facilitate the provision of experimentation services on top of shared testbed(s) and mostly neglected the important aspect of privacy. IoT Data marketplace platforms such as Big-IoT~\cite{bigiot17}, Inter-IoT~\cite{interiot17} and Wise-IoT~\cite{wiseiot16} do provide interoperability but do not focus on data privacy. A federation platform such as FIESTA-IoT~\cite{Agarwal2018FiestaAccess}, although is semantically enabled and provides a module to define user authorisation, does not consider data privacy. However, its semantic power enables IoT testbeds from different domains to federate their data easily, test end-to-end services, collect feedback on performance, and provide experimenters easy to use one-stop data marketplace APIs. 

The lack of focus on data privacy from IoT federation and experimentation platforms motivates us to propose a privacy-enhanced ontology specifically designed for IoT experimentation (that can be easily generalised for generic IoT systems), which includes the main concepts of privacy (as extracted by GDPR and ISO/IEC 29100~\cite{iso2911}) and helps achieve interoperability between testbeds by targeting  horizontal  silos  of  IoT.
{Privacy in IoT experimentation projects is usually addressed using signed agreements with the users, getting their consent to gather their data and use them for experimentation purposes. However, this may cause issues when a user wants to change or revoke her consent, which requires the communication with the testbed owners for requesting these changes. This is usually a timely and manual process, during which the users' data are still getting gathered, stored and (re-)used. Our motivation is to provide an ontology that allows users to dynamically alter their policies and consent in near ``real-time", so that any changes can take immediate effect. Also, the users could create very advanced and personalised policies in this respect. Additionally, the testbed owners will be able to provide new applications, having a process to immediately request the consent of users.} The key contributions of this work are as follows:

\begin{itemize}[leftmargin=*]
    \item \textbf{Ontology alignment}: We follow the methodology proposed by Noy et al.~\cite{Noy2001} and align our ontology to use standard and most popular ontologies such as \textit{SSN}\footnote{SSN: \nolink{\url{http://www.w3.org/ns/ssn/}}\label{f:ssn}} so that we can easily achieve semantic interoperability between semantically enabled IoT testbeds/data-providers.
    \item \textbf{Addition of privacy-related concepts}: We include essential concepts, as per GDPR and ISO guidelines, required for enabling privacy-aware experimentation.
    \item \textbf{Ontology usage recommendation}: We provide recommendations for testbeds/data providers and federation platforms on how to efficiently use our proposed ontology so that the stored data can be utilized. We further provide application scenarios where our ontology fits best.
\end{itemize}

{Note that we do not use the ontologies in totality, rather use and import only those concepts and properties that are needed to support our requirements. Such aspects without which would \textit{(a)} make the ontology model bulky where most of the concepts and properties would be useless and potentially be replicated, \textit{(b)} one testbed might use one concept coming from one ontology while another testbed might use different concept (with same meaning) coming from another ontology, thereby loosing the semantic interoperability.}

{Additionally, the proposed ontology does not intend to cover all GDPR requirements with respect to the full data cycle (data gathering, sharing, processing, retention, deletion, etc.). Our main focus is on the data gathering and sharing aspects, providing entities that cover the phases of when (or not) the data of a user should be gathered and to which other users/applications these data should be shared. This is described in more detail in Section~\ref{sec:req} and Table~\ref{tab:table1}, which shows how our ontology addresses the GDPR requirements. Data retention, processing and deletion are not directly covered by our ontology, but could be indirectly addressed via the privacy policies, if the system is properly implemented (i.e. by having a process to delete data according to a user policy).}

The rest of the paper is organised as follows. Section~\ref{sec:sota} details the related work in the domain of IoT privacy enabled ontologies. Section~\ref{sec:req} describes IoT privacy requirements, while Section~\ref{sec:ontology} and Section~\ref{sec:bp} reports, based on the requirements, the overall in-depth description of our proposed ontology and our recommendations to data provides, experimenters, and platform developers, respectively. Section~\ref{sec:annotate} provide sample workflow, annotations and queries based on the recommendations. Section~\ref{sec:evaluation} presents validation of our ontology based on standard ontology validation tools. We then present use cases and application scenarios in Section~\ref{sec:useCase}. We finally conclude in Section~\ref{sec:conclusion} and summarises our results.

\section{Related work}\label{sec:sota}

An ontology that targets horizontal silos of IoT must address 4W1H (What, When, Where, Who, How) related competency questions and aspects~\cite{Agarwal20184w1h}. Although most of the IoT ontologies  address \textit{what}, \textit{where}, and \textit{when} related competency questions, they fail to address the \textit{who} aspect. In~\cite{Agarwal20184w1h}, the authors survey several ontologies that can be instrumental in addressing 4W1H aspects. They identify that no ontology is comprehensive enough in the IoT domain that can address 4W1H. Popular ontologies such as \textit{SSN}\footref{f:ssn}, \textit{IoT-lite}\footnote{IoT-lite: \nolink{\url{http://purl.oclc.org/NET/UNIS/fiware/iot-lite\#}}, Note, a new version of IoT-lite has a new namespace
\label{f:iotlite}}, and \textit{oneM2M}\footnote{OneM2M: \nolink{\url{https://tinyurl.com/tmnn2qy}}\label{f:onem2m}} all lack concepts, especially related to privacy and access control. \textit{SSN} has concepts to support different IoT systems, and is used by many IoT enabled systems, but although has been refactored it still lacks concepts, such as \textit{service}, \textit{unit of measurement}, and \textit{location}. \textit{OneM2M} provide an abstract model for IoT devices. \textit{IoT-lite} lacks concepts about observations. With many surveys of IoT ontologies already available~\cite{Agarwal20184w1h,agarwal2017study}, we refrain ourselves from going into much detail in this work. Instead, we focus on surveying privacy-related IoT ontologies that have come up recently.

Several ontologies~\cite{abomhara2014security,
sacco2012ppo,costabello2012context,choi2014ontology,imran2016ontology,kirrane2017access,steyskal2014defining, mouliswaran2015inter} have been proposed that target access control and privacy, in general. Developers can reuse privacy and access control concepts from these ontologies, however they do not focus on consent towards accessing particular data, which is a key requirement in GDPR. Fatema et al.~\cite{Fatema2017} proposed a consent ontology targeting the competency question \textit{who is allowed or denied activity on the data}. In another similar work, Pandit et al.~\cite{Pandit2017} proposed provenance and consent ontology. These ontologies are very early work and require much attention to be complete, concise and consistent.

Much recently, \textit{IoT-Priv} has been proposed~\cite{Arruda2019Privacy} extending \textit{IoT-lite} ontology to include \textit{who} aspects and also target privacy. As the core of \textit{IoT-Priv} is \textit{IoT-lite} ontology, \textit{IoT-Priv} focuses on resource storage and access policies. It neglects observations taken by the resources. Moreover, there is no notion of a \textit{data controller}. It also does not follow the conventions proposed by GDPR. \textit{LIoPy}~\cite{loukil2018liopy} and \textit{PrOnto}~\cite{pronto18} both propose privacy-related ontologies that have a legal focus. They do not optimise the ontologies for advanced access control or consider the dynamic context of an IoT environment.

Many EU projects focus on creating frameworks and architectures that introduce privacy concepts in the IoT, but to the best of our knowledge, they do not create ontologies that include privacy features. For example, IoT-A~\cite{bassi2013enabling} developed an ontology as a basis for their architecture (the IoT-A model), but this only includes the main concepts of IoT, without any privacy features. RERUM~\cite{moldovan2016iot} improved the ontology of IoT-A introducing privacy features, but in a very abstract way, since their goal was not to develop an IoT ontology. 

To best of our knowledge, no ontology enables privacy, interoperability, federation, and experimentation in IoT. The closest one being the ontology\footnote{FIESTA-IoT: \nolink{\url{http://purl.org/iot/ontology/fiesta-iot\#}}} defined in~\cite{Agarwal2016unified} (used in FIESTA-IoT platform). This ontology uses concepts from previously well defined ontologies and  defines a taxonomy for the different sensor/actuator devices, phenomena, and measurement units adopted across multiple IoT domains. Nonetheless, it misses the privacy-related concepts.

\vspace{-0.1cm}
\section{Privacy Requirements}\label{sec:req}

Privacy requirements are usually extracted from the eleven privacy principles proposed in ISO 29100~\cite{iso2911}. For IoT, there have been several attempts to adapt the ISO requirements. Here for our ontology, we adopt the eight privacy requirements as defined by the EU project RERUM~\cite{rerumd22,rerumd32}, as an adaptation of the ISO principles combined with the requirements of GDPR. These eight requirements are:
\begin{enumerate}[leftmargin=*]
    \item \textbf{Consent and choice}: IoT applications should only collect data when a user has given explicit consent. The applications should also provide the user with a choice to not allow the collection of her personal information and be able to withdraw her consent at a later stage.
    
    \item \textbf{Purpose legitimacy and specification}: The collection and processing of a user's data should only be allowed when the user provides a specific and legitimate purpose for this collection/processing. Post-processing of collected data for a different purpose than the one for which the data was collected should be forbidden. 
    
    \item \textbf{Collection limitation}: the data that are collected should be adequate and relevant for the purpose and not excessive, with also a limitation for the number of sources from which the data can be collected.
    
    \item \textbf{Data minimisation}: this is closely related to collection limitation and requires that only the minimum amount of necessary data should be collected and processed.
    
    \item \textbf{Accuracy and quality}: data that are collected must be accurate and up to date and IoT services should delete or rectify false/incorrect data.
    
    \item \textbf{Notice and access}: users should be given notice when their data are collected and they should also be able to access their data and delete them when they wish.
    
    \item \textbf{Individual participation and transparency}: users should have full control over their data, knowing who has accessed them in the past and who in general has access to them. Users should also have the ability to start/stop the data collection process at any given time.
    
    \item \textbf{Accountability}: this ensures that whenever there is a privacy breach, the IoT system should be able to hold the responsible person accountable for that breach.
\end{enumerate}


\begin{figure*}
    \hspace{-.5cm}
    \includegraphics[width=1.05\textwidth]{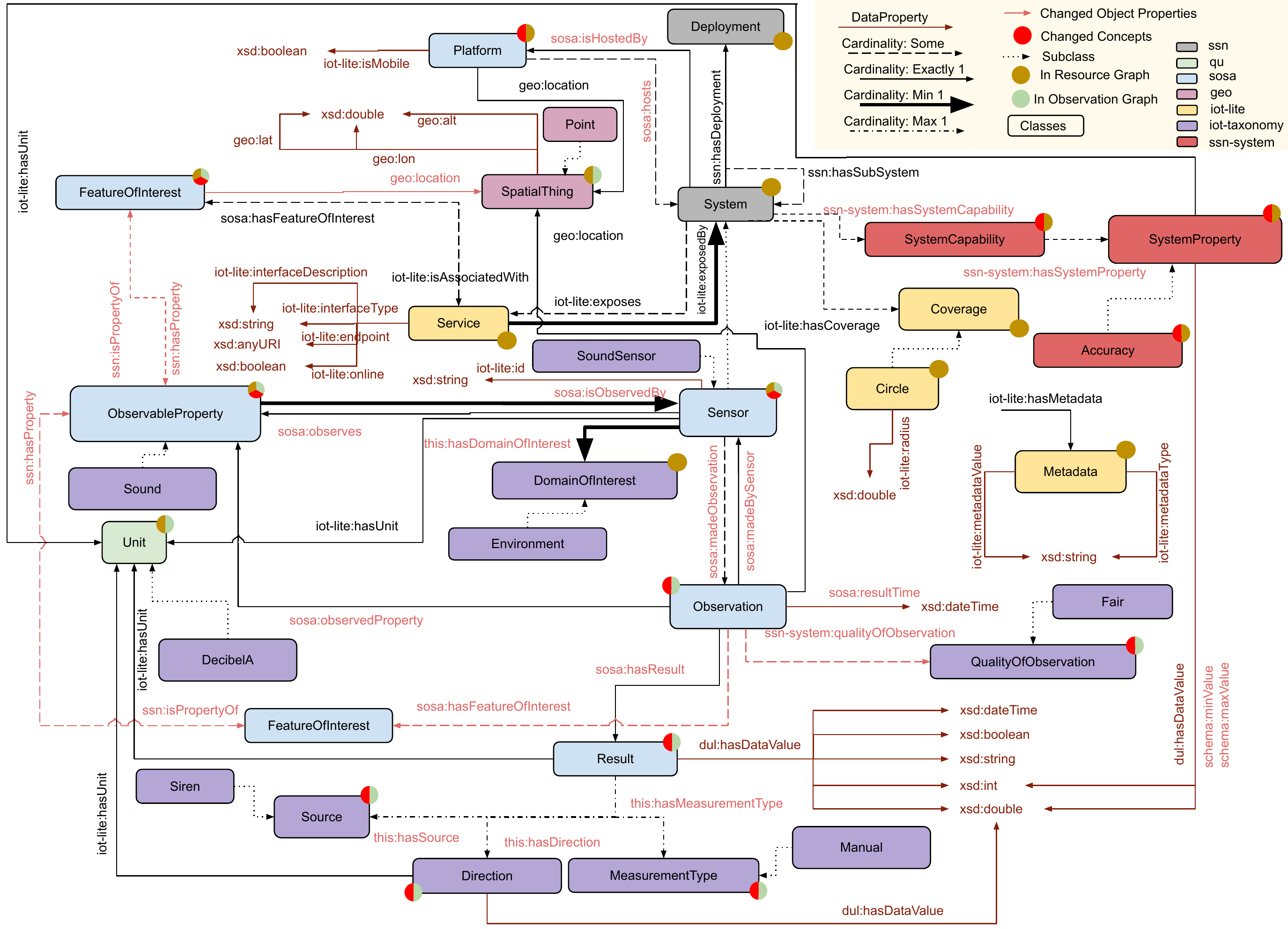}
    \caption{Modified version of Ontology defined in~\cite{Agarwal2016unified}. For clarity we only show concepts related to Sensors. Note that \textit{sosa:Featu-reOfInterest} is replicated just to clearly show the connections. They should be considered same.}
    \label{fig:ontology}
   \vspace{-0.2cm} 
\end{figure*}

\vspace{-0.2cm}
\section{Ontology and related Description} \label{sec:ontology}

Following the methodology in~\cite{Noy2001}, we update the ontology defined in~\cite{Agarwal2016unified} to include concepts enabling interoperability, federation, and experimentation and comply with the new version of \textit{SSN} ontology. {We update the ontology in~\cite{Agarwal2016unified} with a motivation to \textit{(1)} ease migration of testbeds that used ontology in~\cite{Agarwal2016unified} to our ontology version, \textit{(2)} ease new semantically enabled testbed and those that use other standard ontologies to federate with other testbeds.
The ontology in~\cite{Agarwal2016unified} is modelled using two subgraphs (i) a sub-graph that contain essential concepts towards defining a resource (\textit{\textbf{resource graph}}), and (ii) a sub-graph that define the observations taken by a resource (\textit{\textbf{observation graph}}). Similar to~\cite{Agarwal2016unified}, we adopt} all the resource and observation-based concepts and properties in the new proposed ontology {as well} from well-known ontologies such as \textit{SSN}\footref{f:ssn}, \textit{SSN-Systems}\footnote{SSN-Systems: \nolink{\url{http://www.w3.org/ns/ssn/systems}}}, \textit{SOSA}\footnote{SOSA: \nolink{\url{http://www.w3.org/ns/sosa/}}}, \textit{IoT-Lite}\footref{f:iotlite}, \textit{M3-lite}\footnote{M3-lite: \nolink{\url{http://purl.org/iot/vocab/m3-lite\#}}}, \textit{QU}\footnote{QU: \nolink{\url{http://purl.org/NET/ssnx/qu/qu\#}}}, \textit{Geo}\footnote{Geo: \nolink{\url{http://www.w3.org/2003/01/geo/wgs84\_pos\#}}}, \textit{Geo-sparql}\footnote{Geo-sparql: \nolink{\url{http://www.opengis.net/ont/sf\#}}}, \textit{Schema.org}\footnote{Schema.org: \nolink{\url{https://schema.org}}}, and DUL\footnote{DUL: \nolink{\url{http://www.loa.istc.cnr.it/ontologies/DUL.owl\#}}}. Note that, we use \textbf{\textit{owl:equivalentClass}} to link with popular ontologies such as \textit{oneM2M}\footref{f:onem2m}. This enables a testbed that is already semantically enabled to federate with those that would use our ontology. Figure~\ref{fig:ontology} shows the proposed changes in detail. The specific concepts, we used, are colour encoded. A unique colour depicts the ontology from which a particular concept is used. Nonetheless, we mark the changes in the concepts with a red dot and the changes in the properties with a red colour. We kept the core idea of reusing concepts intact and tried to perform minimal yet required changes. 


In Figure~\ref{fig:ontology}, concepts belonging to resource graph are marked with brown dot while those belonging to the observation graph are marked with a blue dot. A multi-coloured dot depicts that a particular concept belongs to more than one subgraph. The changes worth highlighting are:

\begin{itemize}[leftmargin=*]
    \item We update \textit{M3-lite} to reflect the new version of \textit{SSN}. We also move object and data properties defined in \textit{m3-lite} to our ontology and use the \textit{\textbf{this}} prefix to denote them. This makes \textit{M3-lite} clean. We also rename it to \textit{\textbf{iot-taxonomy-lite}}\footnote{iot-taxonomy-lite: \nolink{\url{http://purl.org/iot/vocab/iot-taxonomy-lite\#}}} so to account for new subclass that we created and use  \textbf{\textit{iot-taxonomy}} prefix to represent it.
    
    \item \textbf{\textit{iot-taxonomy:QualityOfObservation}}: the quality of a sensor determines the quality of its observations taken over a period of time. The quality may be low due to calibration issues, but once resolved, the quality can be high. The \textit{iot-taxonomy:QualityOfObservation} identifies the same. To be explicit, in the current version, we define subclasses of \textit{iot-taxonomy:QualityOfObservation} as \textbf{\textit{iot-taxonomy:Poor}}, \textbf{\textit{iot-taxonomy:Fair}}, and \textbf{\textit{iot-taxonomy:Good}}. Although there are other ways of representing the quality of observation, i.e., using \textit{QUDT}\footnote{QUDT: \nolink{\url{http://qudt.org/1.1/schema/qudt\#}}} ontology, but we refrain from using it for this version of our ontology and keep it simple annotation-wise. 
    
    \item We use \textbf{\textit{sosa:Actuator}} to define an Actuating device. Previously, it was defined using \textit{iot-lite:ActuatingDevice}. We also add \textbf{\textit{sosa:Actuation}} and \textbf{\textit{sosa:ActuatableProperty}} to account for the measurement made using \textit{sosa:Actuator}s and the phenomena they act on, respectively.
    
    \item \textbf{\textit{sosa:FeatureOfInterest}}: defines ``\textit{the thing whose property is being estimated}''. For example \textit{sosa:FeatureOfInterest} can be a \textit{room} and a \textit{building}. For now, we add only a few subclasses of \textit{sosa:FeatureOfInterest} and make them available as a part of \textit{IoT-Taxonomy}. A \textit{sosa:FeatureOfInterest} is also associated with \textit{iot-lite:Service}. 
    
    \item \textbf{\textit{ssn-system:SystemProperty}} is ``\textit{the observable characteristic that represents the System's ability to operate its primary purpose}''. For  \textit{ssn-system:SystemProperty} we reuse its \textit{Accuracy}, \textit{Frequency}, \textit{Latency}, \textit{Precision}, \textit{Resolution}, and \textit{Response time} subclasses. We associate these subclasses with three different data properties: (i) \textbf{\textit{dul:hasDataValue}} - used for those system properties that do not have a range of values, (ii) \textbf{\textit{schema:minValue}} and (iii)  \textbf{\textit{schema:maxValue}} - used for those system properties that do have a range of values. We associate the above-mentioned data properties with range \textit{xsd:int} and \textit{xsd:double}.
    
\end{itemize}

We include privacy related concepts in our ontology to address the requirement of privacy and call this new subgraph as a \textbf{\textit{privacy graph}} (see Figure~\ref{fig:privacy}). For the \textit{privacy graph}, we again follow the seven-step methodology~\cite{Noy2001} where we identify the scope of the ontology and before adding concepts in this graph ask competency questions, such as: 

\begin{itemize}[leftmargin=*]
    \item What resources a user has permission to discover?
    \item Who has permission to discover resources and observations?
    \item Who provides the consent for access?
    \item Who owns a particular resource or observation?
    \item What purpose a user has to discover specific types of data?
    \item Who controls the data collection process?
\end{itemize}
    
Given such competency questions, we identified the need to include consent~\cite{Fatema2017} and provenance~\cite{Pandit2017} based concepts. Therefore, we reuse concepts from \textit{Consent}\footnote{Consent: \nolink{\url{http://purl.org/adaptcentre/openscience/ontologies/consent\#}}} (we use prefix \textbf{\textit{con}}) and \textit{GDPRtEXT}\footnote{GDPRtEXT: \nolink{\url{https://w3id.org/GDPRtEXT\#}}} (we use prefix \textbf{\textit{gdprtext}}) ontologies. Nonetheless, these ontologies do not fulfil all the requirements and are not standardised. Upon exhaustive search, we were unable to identify ontologies that we could reuse to fulfil all of our requirements as described in Section~\ref{sec:req}. Thus, we define several object properties to match our requirements. We use prefix \textbf{\textit{priv}} to denote the properties that we have explicitly created. {We model the privacy graph using two subgraphs: \textbf{\textit{consentGraph}} and \textbf{\textit{userPermissionsGraph}}. The \textit{userPermissionsGraph} stores the user permissions for the discovery of either resources or observations along with the permissible actions and their purpose. On the other hand, the \textit{consentGraph} contains information about who has given the permissions to the users and who is the owner of the resources/observations. This graph also stores who controls the data (usually a testbed owner or a user). These privacy related subgraphs are depicted in Figure~\ref{fig:privacy}. The concepts related to consentGraph are marked with a brown dot while those in the \textit{userPermissionGraph} are marked with blue dot. Concepts belonging to both graphs are appropriately shown.}
Next, we describe the concepts and properties we used in the \textit{privacy graph}
, in detail.

\begin{itemize}[leftmargin=*]
    \item The \textit{privacy graph} focuses on the discovery of the resources (\textbf{\textit{ssn:System}} and \textbf{\textit{iot-lite:Service}}) and the observations (\textbf{\textit{sosa:Observation}} and \textbf{\textit{sosa:Actuation}}). Based on the competency questions, we define concepts and properties related to requested permissions, parties that are allowed access, and parties that provide consent to certain permissions. Our \textit{privacy graph} does not focus on the storage of the information about how a permission for discovery is requested, granted, or denied. It also does not store information about the denied parties. We explicitly do this to keep the graph simple.
    

    \item \textbf{\textit{con:AllowedParty}} is the root of the \textit{userPermissionsGraph} and identifies users that are allowed to discover  resources and the observations. \textit{con:AllowedParty} has permission (\textbf{\textit{priv:hasPermission}}) to perform a discovery task. Thus, \textit{priv:hasPermission} has the domain \textit{con:AllowedParty} and the range \textbf{\textit{con:Permission}}. It has an inverse object property called \textbf{\textit{con:permission\_given\_to}}. The \textit{con:AllowedParty} can have no or more than one \textit{con:Permission}s. Similarly, a \textit{con:Permission} is given to no or more than one \textit{con:AllowedParty}. 
    
    \begin{figure}
    \centering
    \includegraphics[width=\columnwidth]{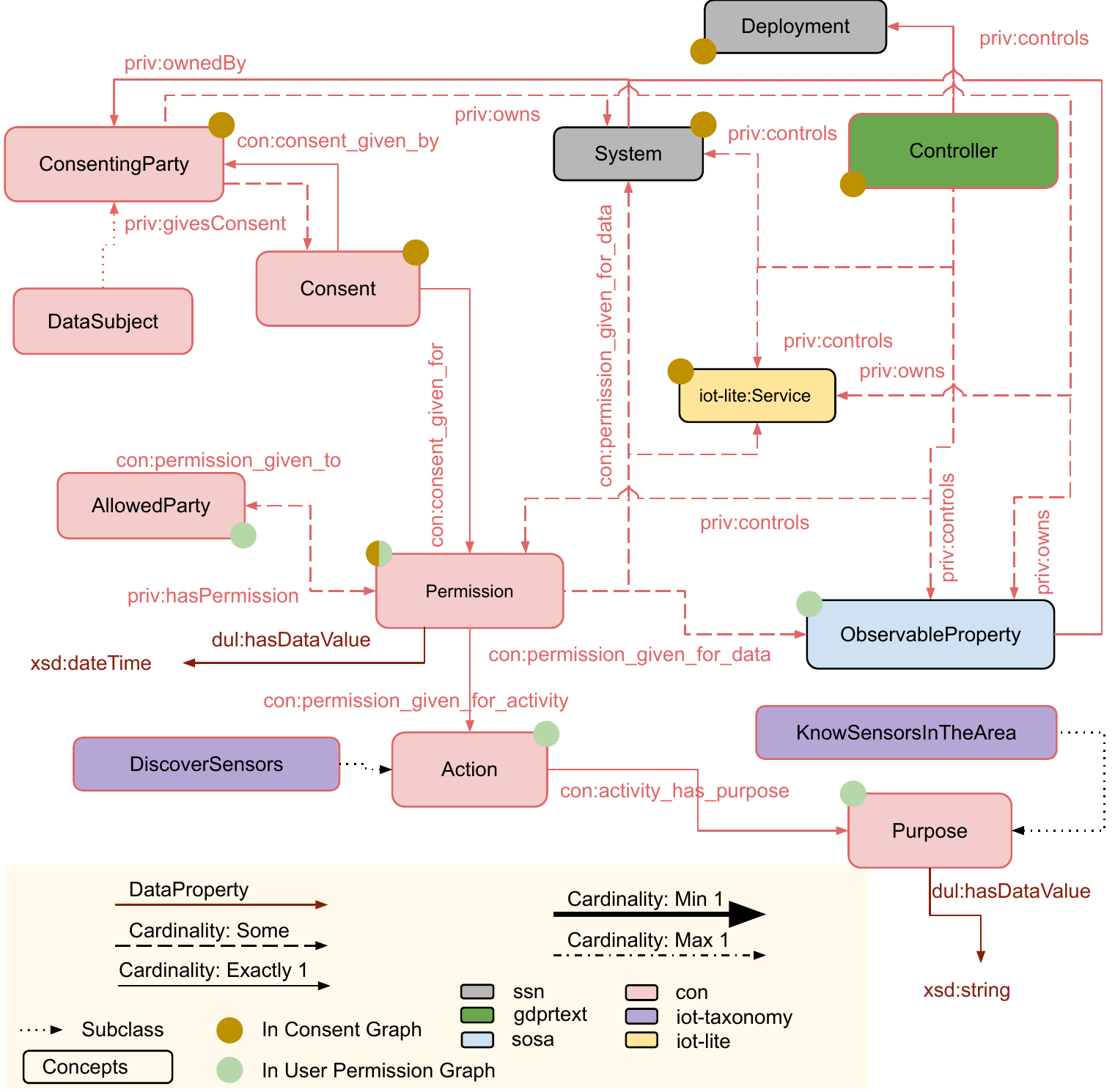}
    \caption{Privacy related concepts included in the ontology.}
    \label{fig:privacy}
    \vspace{-0.4cm}
\end{figure}
    
    \item \textbf{\textit{con:Permission}} identifies different permissions in the system. Permissions are given for (\textbf{\textit{con:permission\_giv-en\_for\_data}}) discovering the resources (\textit{ssn:system} and \textit{iot-lite:service}) and for phenomenon (\textit{sosa:ObservedProperty} and \textit{sosa:ActuatableProperty}). Some \textit{con:Permission}s can be denied or given to discover the above data. Note that here we specifically link to \textit{sosa:ObservedProperty} and \textit{sosa:ActuatableProperty} to avoid any scalability issues that would appear in case we link to \textit{sosa:Observation} or \textit{sosa:Actuation} because different observations would be stored in the triple-store as and when they are made. Using the link between \textit{sosa:observableProperty} and \textit{sosa:Sensor} users can query the graph to get \textit{sosa:Observations}.
    
    \item \textbf{\textit{con:Permission}} is given for a particular time for a particular \textbf{\textit{con:Action}} which, for example, can be the discovery of sensors (\textbf{\textit{iot-taxonomy:DiscoverSensors}}). The \textbf{\textit{dul:hasDataValue}} identifies the time until when the permission is granted. We assume that when an entry is made in the triple-store, permission is granted. The object property \textbf{\textit{con:permission\_given\_for\_activity}} that has domain \textit{con:Permission} and range \textit{con:Action} helps to identify the set of activities that are granted to a \textbf{\textit{con:AllowedParty}} and for what purpose (\textbf{\textit{con:Purpose}}). Each \textit{con:Action} has a purpose. \textbf{\textit{con:activity\_has\_purpose}} that has domain \textit{con:Action} and the range \textit{con:Purpose} helps provide such information. A \textit{con:Purpose} can be either free text (\textbf{\textit{xsd:string}}) or from a predefined set (example: \textbf{\textit{iot-taxonomy:KnowSensorsInTheArea}}).
    
    \item The \textit{consentGraph} contains information about who is permitted to access the resources and the observations. The root of \textit{consentGraph} is the \textbf{\textit{con:ConsentingParty}} which identifies \textit{the person who has given the consent} (\textbf{\textit{con:Consent}}) and also owns (\textbf{\textit{priv:owns}}) the resources and the observations. \textit{priv:owns} has an inverse property called \textbf{\textit{priv:ownedBy}}. The \textit{con:ConsentingParty} can \textit{priv:owns} multiple resources and observations while a resource or an observation is \textit{priv:ownedBy} by exactly one \textit{con:Consenting-Party}. \textbf{\textit{con:gives\_consent}} links \textit{con:ConsentingParty} and the \textit{con:Consent} where the domain is \textit{con:ConsentingParty}, the range is \textbf{\textit{con:Consent}} and the cardinality is 0 or more.
    
    \item \textbf{\textit{gdprtext:Controller}} is the ``\textit{legal person, public authority, agency or other body which, alone or jointly with others, determines the purpose and means of the processing of personal data.}'' It is the administrator of the testbed who controls the testbed and the related data.
    
\end{itemize}
Our ontology addresses all the requirements {that are related with data capturing and sharing phases} described in Section~\ref{sec:req}. A mapping is presented in Table~\ref{tab:table1}. 

\begin{table}
    \caption{Requirements mapped to our ontology}\label{tab:table1}
    \begin{tabular}{c|p{7.90cm}}
        \hline
        \textbf{\#} & \textbf{How the requirement is addressed in the ontology} \\
        \hline
        1 & Given a specific \textit{purpose}, there are entities for \textit{consent} and \textit{consenting party} that give/revoke permission to the action (i.e., data collection) requested by some users. \\
        \hline
        2 & The entity \textit{purpose} is introduced to require that access to data is given only for a specific purpose and not in general\\ \hline
        3 & {Compliance} to the ontology requires that any \textit{action} is given permission only for a specific \textit{purpose} and the privacy policy specifies which data should be collected and that only these are collected. \\  \hline
        4 & Similarly, as above, the entity \textit{action} (for a specific  \textit{Purpose}) controls the type and amount of data that are requested for a service according to a specific purpose and then grants permission for the minimum amount of data to be collected.\\ \hline
        5 & Users can know the quality of the data stored and this can be done using the \textit{accuracy} entity which allows accuracy values to be sent together with the observations.\\ \hline
        6 & The \textit{consent} module supports the notification of the users when there is a new request for collecting their data. \\ \hline
        7 & Users can be considered as \textit{allowedParty} to have access to their own data and know at any given time what data are collected and who has access to their data. This can be implemented as an additional IoT Service on top of the IoT System, with users having additional, administrator privileges on their privacy policies.\\  \hline
        8 & An IoT system should keep record of users that have requested access to data and can backtrack to identify the origin of a data breach when it takes place. The \textit{AllowedParty} entity can help on this aspect.\\ 
        \hline
    \end{tabular}
    \vspace{-.2cm}
\end{table}
\vspace{-.1cm}


\section{Best practices followed and recommendations to triple-store creators and data providers}\label{sec:bp} 

\begin{figure}
    \centering
    \subfloat[Resource Graph Annotation][Resource Graph Annotation]{
        \includegraphics[width=\columnwidth]{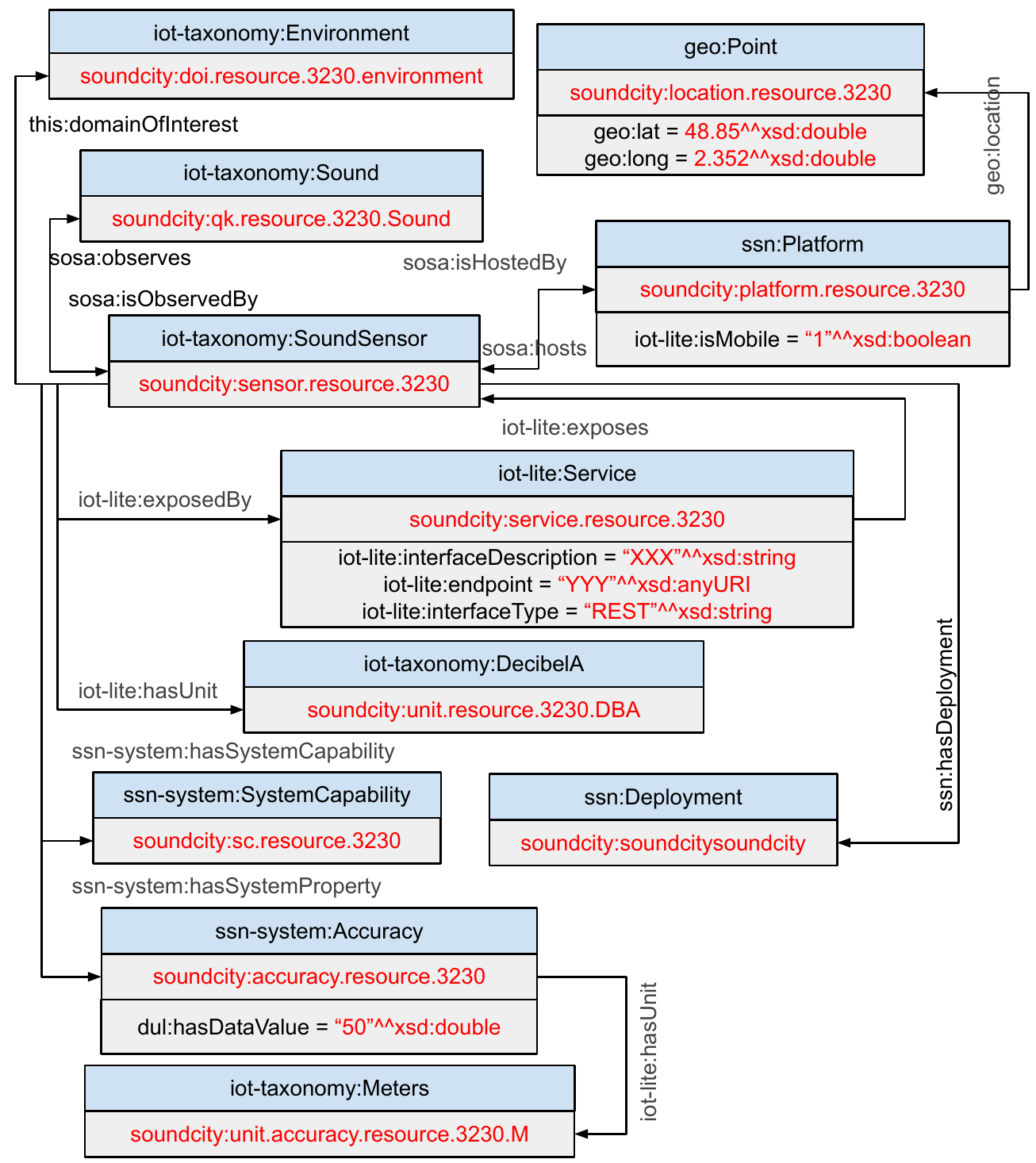}
        \label{fig:resourceannotation}
    }\\
    \subfloat[Observation Graph Annotation][Observation Graph Annotation]{
         \includegraphics[width=0.9\columnwidth]{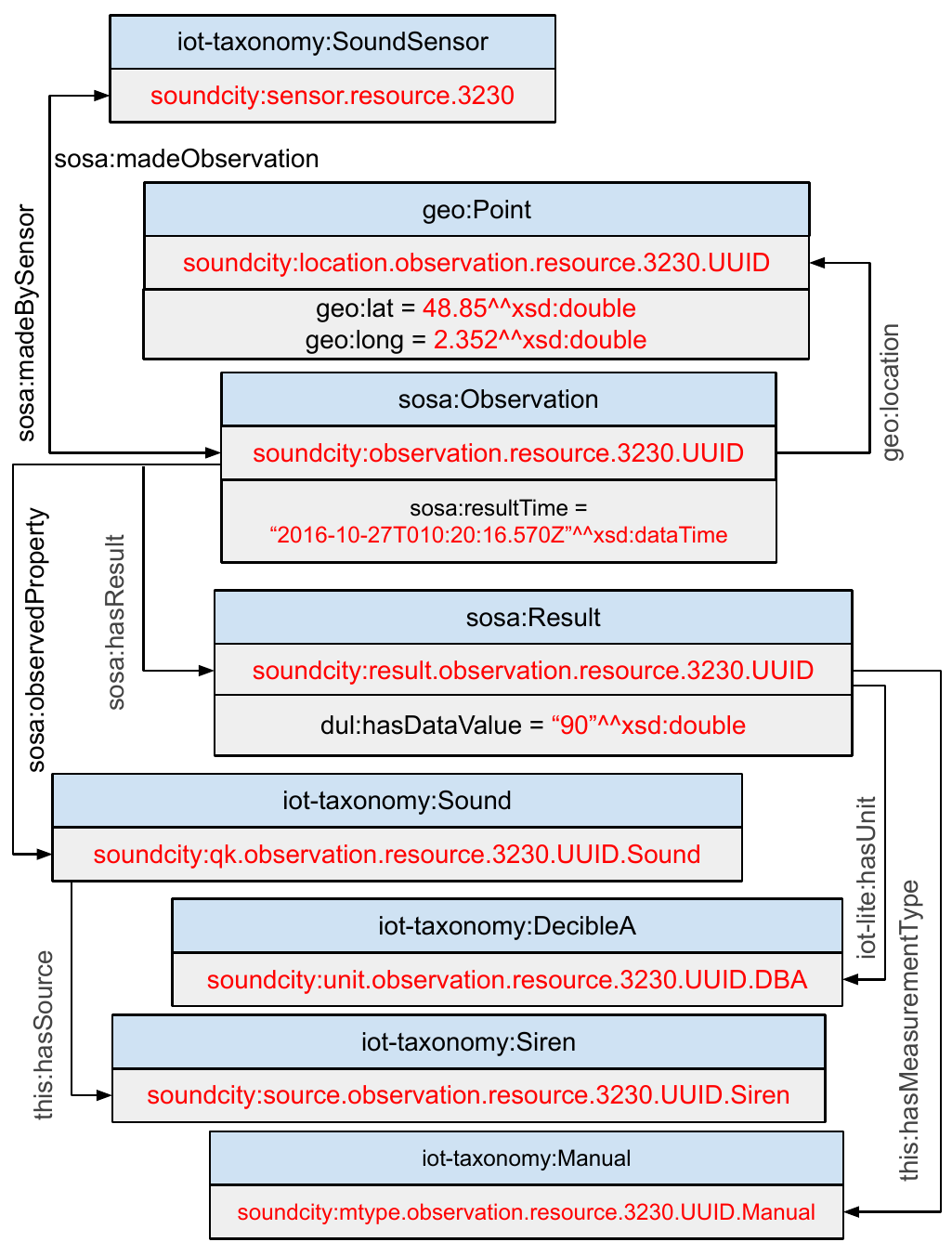}
        \label{fig:observationannotation}
    }
    \caption{Sample annotations for resource and observation graph.}
\vspace{-.2cm}
\end{figure}

For an ontology to have impact, be re-usable, and interoperable it is necessary that it follows best practices and the data stored using the ontology follow recommended guidelines. Below we list the steps we followed to {build} our ontology and our recommendations to testbeds and triple-stores that use {it}.
\vspace{-.12cm}
\subsection{Best practice followed to create the ontology}
Our ontology follows the methodology described in~\cite{Noy2001}. For IoT systems on the top of the said methodology, the ontology is engineered to follow recommendations described in~\cite{Gyrard2015}. Most of the concepts and properties, other than those that we have defined, are taken from ontologies that already follow the best practices and are available in LOV\footnote{LOV: \nolink{\url{https://lov.linkeddata.es/dataset/lov/}}}~\cite{vandenbussche2017linked}. Further:

\begin{itemize}[leftmargin=*]
    \item We publish the ontology in several formats, such as \textit{RDF/XML}, \textit{JSON-LD}, \textit{Turtle}, and \textit{Ntriples}. Each concept and property fulfills a set of metadata that includes \textit{rdfs:labels} and \textit{rdfs:comments}. We maintain the online accessibility\footnote{FIESTA-Priv: \nolink{\url{http://purl.org/iot/ontology/fiesta-priv\#}}} and the permanent URL of the ontology. The ontology web documentation is generated using WIDOCO~\cite{Garijo2017}. It reflects detailed documentation of the concepts and the properties, how-to, and related material.
    
    \item We aim to standardise the ontology. As a step towards standardisation, the ontology is in submission to LOV.
    
    \item In case any testbed/data provider wants to request updates in our ontology or taxonomy towards integration in the federated triple-store, they can use our GitHub issue tracker\footnote{FIESTA-Priv Github: \nolink{\url{https://github.com/ragarwa2/fiesta-priv}}} to propose changes. For completeness, each concept or property that the testbed wants to add should consist of a name (\textit{rdfs:label}), a clear description (\textit{rdfs:comment}), and the parent (if any). Further, for properties, additionally, the testbeds should define the domain, the range, and the cardinality. To ease the integration process, if the concept or the property is already defined in some other ontology, the testbed/data provider should also provide the \textbf{\textit{rdfs:isDefinedBy}} or \textbf{\textit{rdfs:seeAlso}}. 

\end{itemize}
\vspace{-.12cm}
\subsection{Recommendations to federating triple-stores and testbed/ data providers} As discussed before, there are four subgraphs in our ontology. The interactions of the \textit{consentGraph} with the resource graph and the  observation graph increase with the number of resources and observations. This increases communication and storage overheads. Thus, {we provide recommendations that are particular to our ontology. These recommendations certainly should be treated along with the generic recommendations that exists within the semantic realm. Our recommendations are:}
    \begin{itemize}[leftmargin=*]
        \item { related to the use of privacy graph.}
        \item { related to the use of \textit{iot-lite:metadata}.}
        \item { related to the order in which data should be inserted.}
    \end{itemize}
    
    {The following list provides our specific recommendations:}
    
    \begin{figure}
    \centering
    \subfloat[Consent Graph Annotation][Consent Graph Annotation]{
        \includegraphics[width=0.85\columnwidth]{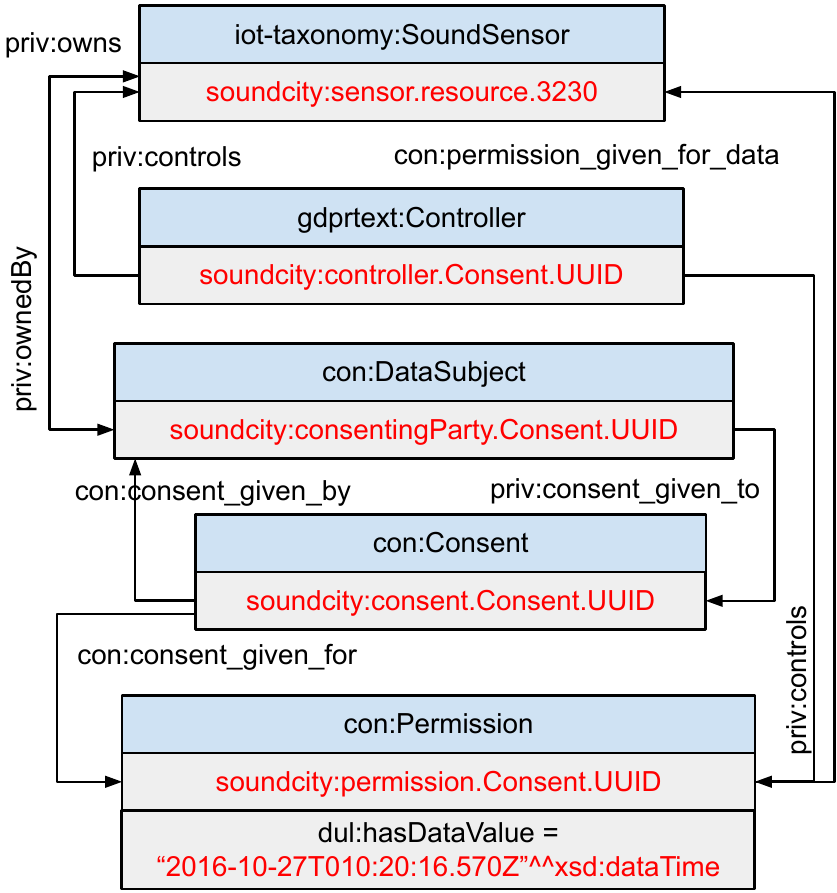}
        \label{fig:annotationConGraph}
    }\\
    \subfloat[UserPermission Graph Annotation][UserPermission Graph Annotation]{
        \includegraphics[width=0.85\columnwidth]{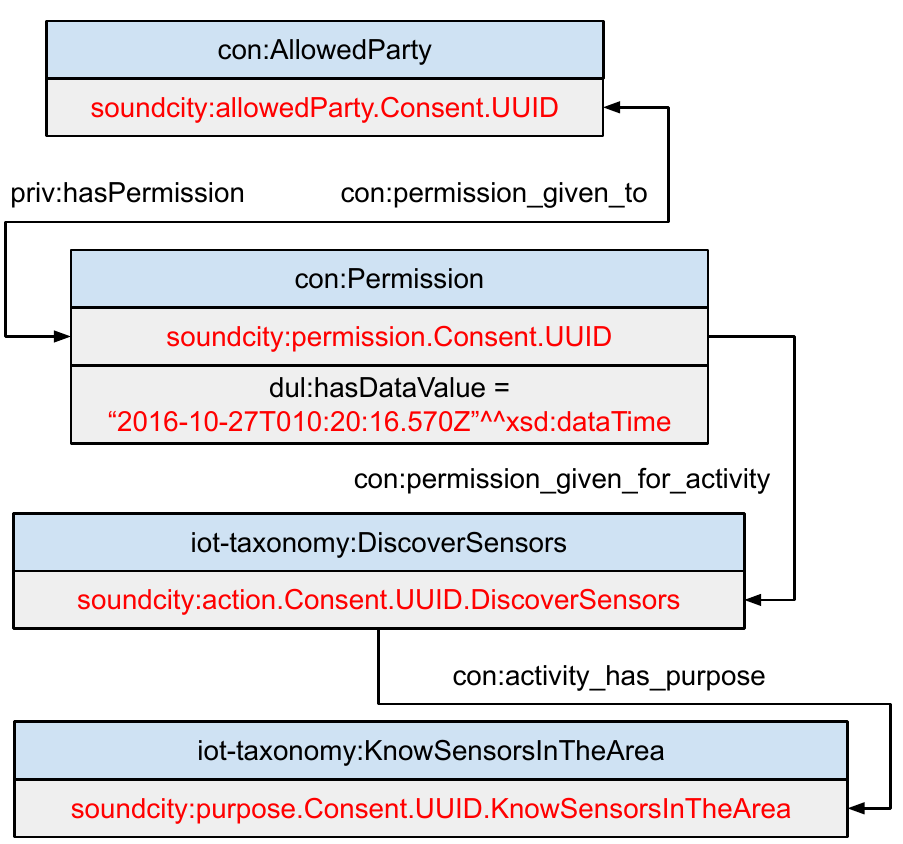}
        \label{fig:annotationAllowedParty}
    }
    \caption{Sample annotations for privacy graph.}
\vspace{-.25cm}
\end{figure}

    \begin{itemize}[leftmargin=*]
    \item {Using our ontology, currently, one cannot store information regarding how the permissions are granted or denied. It focuses only on which users are granted permissions. The triple-store should provide the functionality to update the \textit{consentGraph} and the \textit{userPermissionGraph} with relevant information when a new resource or new observation is added. Further, the functionality should also enable the update of the \textit{userPermissionGraph} when a new permission is granted. Nonetheless, when permission is denied, the relevant information should be deleted from the graph using the triple-store provided functionality. Also, our ontology does not store any information about how a user gets registered or any other information about the user. It only contains the URIs of users that are \textit{con:AllowedParty} or \textit{on:ConsentingParty}. A triple-store should build their own user management functionality.}
    
    \item {The privacy graph should have restricted access. No user except the triple-store admin should be allowed to query this graph. \textit{con:AllowedParty} should only have access to those resources and observations for which they have permission (i.e., their own data or policies/permissions).}
    
    \item {A testbed should respect the inverse properties and cardinalities of the object properties. If the inverse property exists, they should provide triples to support both links. Moreover, if a testbed wants to publish triples related to some feature that is not supported currently by the ontology, they can use \textit{iot-lite:Metadata}. Nonetheless, they should not abuse the \textit{iot-lite:Metadata} concept.}
    
    \item {IoT resources should first be inserted into the resource graph. Then privacy related information should be updated. Then testbeds should publish observations in the observation graph. Note that if a resource or \textit{sosa:ObservableProperty} is not granted any permission, then they are not accessible to any \textit{con:AllowedParty} but only the \textit{con:ConsentingParty} and the \textit{gdprtext:Controller} can access them.}
    \end{itemize}

\vspace{-.2cm}
\section{Workflow, Annotations and Querying}\label{sec:annotate}

The workflow for annotation and access to resources and observations would primarily depend on the system design for the platform, but as a guideline, we propose the following approach. At the initial stage the data provider should register resources that it wants to make available on the platform for experimentation and populate the \textit{resource graph}. Post this process, the data provider creates instances for \textit{gdprtext:Controller} and \textit{con:ConsentingParty} using in the \textit{registration} module, unless this is set by default by a prior data provider. User access to any resource or observations belonging to the data provider and residing in the triple-store of the platform is denied by default, unless a policy is pre-set. The policy would primarily be based on the \textit{con:Purpose} and \textit{con:Action}, due to their significance with GDPR compliance. Assume, \textit{con:AllowedParty} are users that are registered. For such a case the data provider creates a \textit{con:Permission} that will link all the current and any future \textit{con:AllowedParty} with a resource if the policy requirements are met. This type of process can be viewed as ``passive'' consent. If a policy is not pre-set then a user (\textit{con:AllowedParty}) request for access to resources or observations is triggering a mechanism to request an ``active" consent usually within a set period before expiration. This would involve the platform notifying all the \textit{con:ConsentingParty}s. Whenever a \textit{con:ConsentingParty} grants access to their resources or observations, a new \textit{con:Consent} and \textit{con:Permission} is instantiated and the usersPermissionGraph is updated accordingly.

For a better understanding of our recommendations and workflow, we provide sample annotations that a testbed should refer. Figure~\ref{fig:resourceannotation} and Figure~\ref{fig:observationannotation} provide sample annotations for the resource and observation graph. The annotations assume that a sound sensor associated to an example \textit{testbed} called \textbf{\textit{Soundcity}} (a mobile crowdsensing testbed~\cite{Agarwal2018Sensors}) is available and  
the testbed wants to use our ontology to both store data locally in its own triple-store and send the data to a federated triple-store. In the process the testbed creates an IRI \textit{soundcity:sensor.resource.3230} of type \textit{iot-taxonomy:soundSensor}. The sensor is on a platform \textit{soundcity:platform.resource.3230} that has location \textit{soundcity:location.resource.3230} and is mobile (\textit{iot-lite:isMobile}=\textit{1\string^\string^xsd:boolean}). The sensor is exposed via a service \textit{soundcity:service.resource.3230} and has unit \textit{soundcity:unit.resource.3230} of type \textit{iot-taxonomy:DecibelA}. The sensor has domain of interest \textit{soundcity:doi.resource.32-30.environment}. The testbed similarly creates IRIs to other associated properties. The testbed annotates an observation produced by this sensor at time \textit{2016-10-27T0-10:20:16.570Z\string^\string^xsd:dataTime} (after setting the required privacy parameters) in a similar way and creates its IRI \textit{soundcity:observation.resource.3230.UUID}. Here UUID is the unique identifier. This observation has \textit{sosa:Result} IRI \textit{soundcity:result.observation.resource.3230.UUID} which \textit{dul:has-DataValue} \textit{90\string^\string^xsd:double}. This observation is recorded towards the sound produced by a sound source \textit{soundcity:-source.observation.resource:3230.UUID.Siren}. Note that, in these figures, the ontology concepts are shown in light blue box while the testbed provided IRIs are shown in light grey box. The figures also show the object properties and the annotations only to the most relevant concepts. Note that the starting point in the resource graphs is \textit{soundcity:sensor.resource.3230} while for the observation graph, depending on the need, it can be either \textit{soundcity:sensor.resource.3230} or the observation \textit{soundcity:observation.resource.3230.UUID} the sensor made. This is in particular useful while querying.

\begin{figure}
\hspace{-1.7cm}
    \includegraphics[width=\columnwidth, , angle =270]{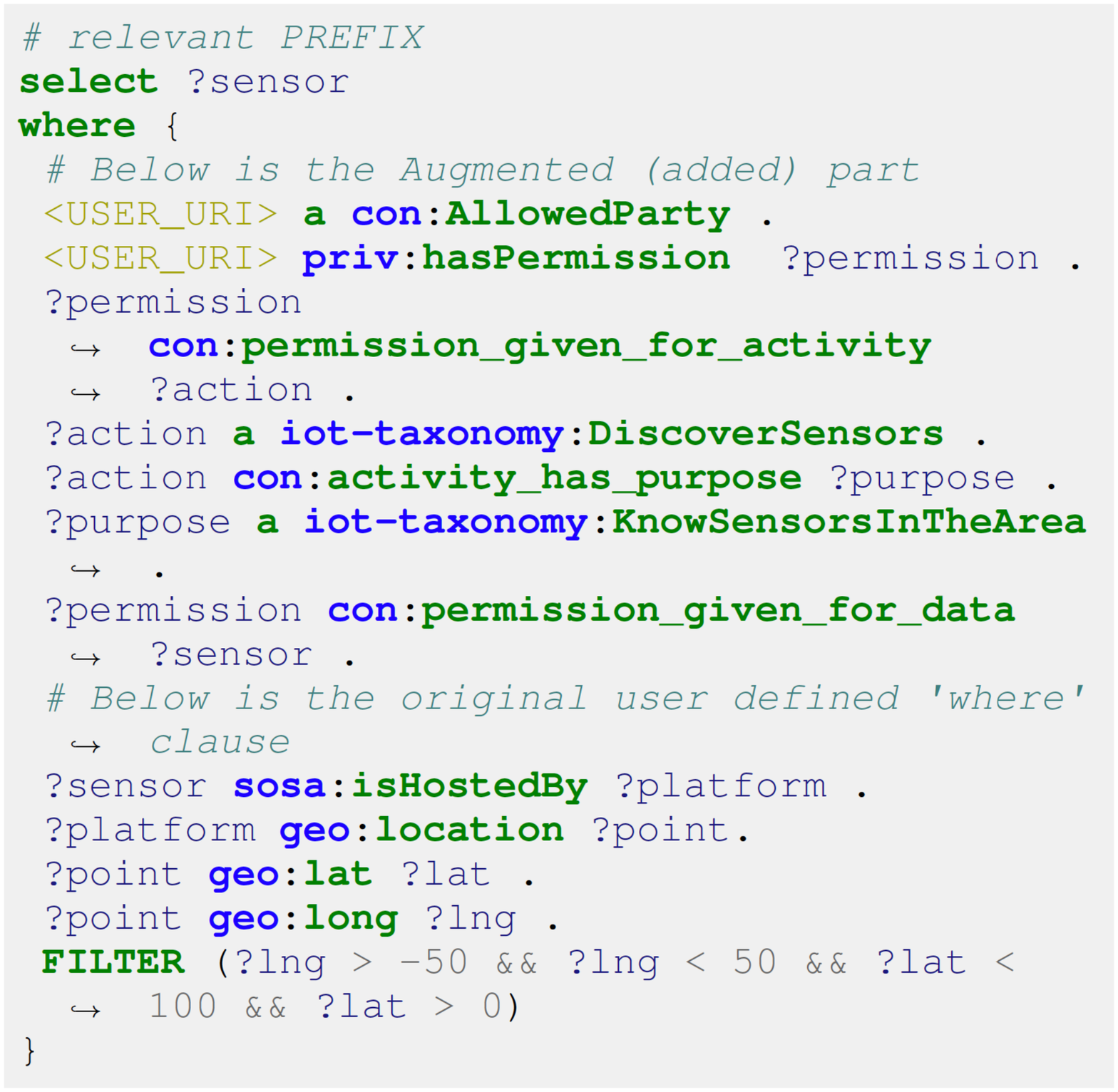}
\caption{Augmented SPARQL query.}
\label{fig:augmentedSparqlQuery}
\vspace{-.45cm}
\end{figure}

On the other hand, Figure~\ref{fig:annotationConGraph} and Figure~\ref{fig:annotationAllowedParty}, provide sample annotation for the \textit{consentGraph} and \textit{userPermissionGraph}, respectively. These figures as before use same styling and show annotations to most relevant privacy related concepts. Here, for the same sensor described above, the testbed provides annotations regarding a consenting party (IRI \textit{soundcity:consentingParty.Consent.UUID}) who is of type \textit{con:DataSubject} that gives consent (\textit{soundcity:consent.Con-sent.UUID}) for a request of permission (\textit{soundcity:permis-sion.Consent.UUID}). Depending on the the scenario, this permission could be asked by the user/experimenter for the activity of discovering sound sensors (\textit{soundcity:activity.-Consent.UUID.DiscoverSensors}) for the purpose of knowing the available sensors in the given area (\textit{soundcity:purpo-se.Consent.UUID.KnowSensorsIntheArea}) or set by the testbed itself (passive mode). Note that, as consentGraph is data provider centric, the starting point or the root of the graph is the \textit{con:ConsentingParty}, \textit{soundcity:consentingParty.Con-sent.UUID}. On the other hand, depending on the needs, for the userPermissionsGraph, the starting point can be either the \textit{con:AllowedParty} or the \textit{con:Permissions}.

Once data is available in a triple-store, users/experimenters can write SPARQL queries depending on their needs using the SPARQL endpoint provided by the triple-store. As the users/experimenters only need to access the resource and observation graphs, they should only write queries that target either the resource, the observation graph or both.
Consider an experimenter who has registered his interest for discovering available sensors (\textit{iot-taxonomy:DiscoverSensors}) in a given location for knowledge purpose (\textit{iot-taxonomy:KnowSensorsIntheArea}) with the platform and the platform has stored this information. The userPermissionsGraph is only updated after a particular interest/request of the user is granted permission. The user then makes a query to get data associated to all the sensors available in a region bounded between -50 degrees and +50 degrees longitude and between 0 degree and 100 degrees latitude.
The triple-store then augments the query internally to include the privacy related concepts depending on the registered interest of the user. On the other hand, the triple-store can also first execute the user query and then filter the results based on the privacy graph (a time-consuming alternative). 

Assuming the triple-stores use the first approach, Figure~\ref{fig:augmentedSparqlQuery} shows an augmented SPARQL query for knowing all resources a particular user can discover in a particular region. 
Here, we assume that the platform internally gathers all the permissions requested by the user and use them in the augmentation part. In the example, as the user wants to \textit{iot-taxonomy:DiscoverSensors}, the platform augments the statement \textit{?action a iot-taxonomy:DiscoverSensors} in the query.

\vspace{-0.2cm}
\section{Ontology Validation}\label{sec:evaluation}
{We perform a qualitative evaluation of the ontology, validating its conformance (example: lexical, syntactic and semantic aspects)~\cite{datta2018} and quality (example: accuracy, completeness, conciseness, adaptability, clarity, computational efficiency, and consistency)~\cite{Raad2015}. We also comment on the reusability and modularity aspects. 
This work does not provide any usage specific evaluation as we do not focus on building a platform.}


For structural aspects, we validate our ontology using tools such as Ontology Pitfall Scanner\footnote{OOPS: \nolink{\url{http://oops.linkeddata.es}}} (OOPS), TripleChecker\footnote{TripleChecker: \nolink{\url{http://graphite.ecs.soton.ac.uk/checker/}}} and Vapour\footnote{Vapour: \nolink{\url{http://linkeddata.uriburner.com:8000/vapour}}}. Among the above tools, OOPS also provides us with an evaluation of completeness, conciseness, clarity, and consistency. An OOPS report on the validation of our ontology is available on the ontology document page. The tool identifies that our ontology is concise. For completeness, it reports that our ontology has unconnected elements. This is true in the case of \textit{iot-lite:Metadata} for which domain is deliberately kept open. It further reports that some properties are missing domain and range. This again is not relevant as we have used \textit{schema:domainIncludes} and \textit{schema:rangeIncludes} to define domain and range of the properties. These annotations are not checked by the tool. The tool also reports that there are inverse relations other than those defined. This again is not true as the tool reported inverse relations are not valid. Thereby making our ontology consistent. The only clarity issue raised is the use of different naming conventions in the ontology. We are aware of this situation as the \textit{Consent} ontology uses a different convention. Other than OOPs report, we also make available the TripleChecker\footnote{FIESTA-Priv TripleChecker: \nolink{\url{https://tinyurl.com/qwbnj87}}} and Vapour\footnote{FIESTA-Priv Vapour: \nolink{\url{https://tinyurl.com/r7f5wbt}}} reports.

To provide complete validation of our ontology, we execute the Hermit reasoner with the hope to identify any entailments that we did not intend. We apply reasoning on the structure of the ontology, the relationships among classes and among properties. As there are no instances, we do not reason instances. For our ontology, other than direct relations, the reasoner inferred results mainly using rules such as: 
\begin{itemize}[leftmargin=*]
        \item (rdfs11) if A is a subClassOf B and B is a subClassOf C then A is a subClassOf C
        \item if A is a subClassOf B and B is equivalentTo  C then A is a subClassOf C
        \item if A is a subClassOf B and B is inverseOf C then A is a subClassOf inverseOf C
        \item if A is inverseOf B and B is equivalentTo C then A is inverseOf C 
\end{itemize}
Note that here we only show the rules for classes. Similar results were obtained for objectProperties as well. Nonetheless, we notice no issues\footnote{The results of Hermit reasoner are available at \nolink{\url{http://smart-ics.ee.surrey.ac.uk/fiesta/ontology/fiesta-priv/HermitResults.owl}}}. The reasoner also detected that our ontology is `satisfiable'. For validation purpose we also executed Pallet Reasoner and found similar results.

Our ontology is partially modular. It reuses most of the concepts from the existing ontologies. We defined the concepts and properties within the ontology because we were not able to find related concepts in the existing ontologies. 

According to the ontology metrics, our ontology has 3133 Axioms, 645 Logical Axioms, 608 Declaration Axioms, 520 Classes, 46 Object Properties, 13 Data Properties, 31 Annotation Properties, 10 Inverse Object Properties, and 3 Equivalent properties. Out of the above-mentioned number of classes, object properties, and data properties, we have only created 10 object properties. All the other classes, object properties, and data properties have been reused from existing ontologies.
    
\vspace{-0.1cm}
\section{Use cases}\label{sec:useCase}
We demonstrate the scenarios where our ontology can be used to ensure privacy for testbed participants and enable testbed providers to restrict sharing their data with the users.
\vspace{-0.1cm}
\subsection{Selective Privacy for Participants}

Testbeds and Living Labs are often closely associated with people. Data captured is often multi-variate in the sense that IoT devices associated with a testbed usually monitor multiple phenomena simultaneously. For example, a testbed may comprise devices installed in a room that measure temperature, humidity, sound, and light intensity. Some participants might not feel comfortable with sharing data about sound that is around them, even though devices are not recording audio streams. Additionally, temperature information may also reveal user presence (having the air-conditioning in \textit{on} state). Thus, the associated testbed must take the user's acceptance, and ultimately consent, into serious consideration and restrict access to these data on the level mandated by the participant. This could be based on user roles, affiliation or locality. In this case, the \textbf{\textit{con:ConsentingParty}} would not give \textbf{\textit{con:Consent}}, and in turn the \textbf{\textit{con:Permission}} to the \textbf{\textit{sosa:ObservableProperty}} would be denied. Based on this, the testbed provider who is in \textbf{\textit{priv:control}} as the \textbf{\textit{gdprtext:Controller}}, will enforce denial of access to consumers who are not an \textbf{\textit{con:AllowedParty}}.

In cases where devices are carried by humans or are mobile, such as mobile crowdsourcing, an \textit{con:Actions} relating to discovering sensors or observations can be restricted based on spatial and/or temporal rules (restricted to certain locations and to certain times of the day). For example, consider a living lab deployed in a workplace where sensory data is captured from employee smartphone managed by an employer. The employee might not want his employer to know what he does or where he goes outside of working hours. Here the employee (\textbf{\textit{con:ConsentingParty}}) can restrict the discovery of \textit{sosa:ObservableProperties} (\textbf{\textit{con:Action}} being \textbf{\textit{iot-taxonomy:GetWorkplaceObservations}}) captured during work hours only to the employer (\textbf{\textit{con:AllowedParty}}).
\vspace{-0.1cm}
\subsection {Selective Data Sharing for Providers}

IoT-enabled platforms have evolved from collecting data from homogeneous to heterogeneous sources. This involves integrating systems from multiple devices where data from associated devices are sent to the testbed where it is stored, analysed, and monitored. In the data-driven world one of the major needs of the experimenters is the volume of the data so as to perform different analysis such as comparison between data of same quantity kind, and to further train their AI components to achieve better performance. Here, a data exchange point can be established by having a central or federated repository to pull data from, or more practically a data broker where experimenters can subscribe to data of their interest. Although, providers/testbeds might want to restrict exposing their data to experimenters based on the purpose (\textit{con:Purpose}) of data consumption, information governance compliance, or no explicitly declared interest submitted by the experimenter. Also, some providers/testbeds might want to deny access to their data (or subset of) to other providers who are identified as direct competitors. Data providers (\textit{con:ConsentingParty}) can only allow \textit{con:AllowedParty} for the data access.  

From the consumer perspective, users with different roles would have access to the data. For example, in the context of healthcare trial systems, these users could be clinical monitoring staff (or first point of contact), general practitioners, society support members, technicians or researchers. End users such as patients or carers participating in a trial can restrict which roles are allowed access to their data.

\vspace{-0.1cm}
\subsection{Applicability in IoT Platforms and Existing IoT Ontologies}\label{sec:application}
Besides testbeds/data providers that have the motivation to federate their data or provide semantic powers to it, our ontology finds applicability on IoT platforms that support federation, semantic interoperability, and experimentation, as well as on different existing well known IoT ontologies.

The ontology defined in~\cite{Agarwal2016unified} has been successfully integrated with IoT Cloud Platforms such as FIESTA-IoT~\cite{Agarwal2018FiestaAccess} and has been used to federate at least 11 testbeds~\cite{Agarwal2018Sensors} that support more than 23 experiments. These testbeds and experiments have different application domains (such as smart cities, crowdsensing, smart agriculture, smart building, smart energy, and smart sea) and were selected based on their usefulness, complementarity, sustainability, and competence. With such a success and data being released to third parties for experimentation purposes, privacy aspects (not addressed in~\cite{Agarwal2016unified}) must be integrated within the FIESTA-IoT Platform so to comply with GDPR. Thus, integrating our ontology with the FIESTA-IoT Platform will provide an edge to FIESTA-IoT. The same applies to other IoT Platforms such as Wise-IoT~\cite{wiseiot16} and ACTIVAGE~\cite{Karaberi2018} that adopt ontology defined in~\cite{Agarwal2016unified}. Such adoption would not only help experimenters that include industries, researchers, and students but also teachers who can use federated data in their courses. 

Several IoT ontologies that lack privacy concepts can adopt the privacy graph as a whole or the object properties defined therein to make themselves privacy enabled. As an example, well know ontologies those missing privacy concepts such as SSN\footref{f:ssn}, IoT-lite\footref{f:iotlite}, OneM2M\footref{f:onem2m} and SAREF\footnote{SAREF: \nolink{\url{http://saref.linkeddata.es}}} including its extensions can leverage from our defined privacy graph.

\vspace{-0.1cm}
\section{Discussion and Conclusion}\label{sec:conclusion}

In the future IoT-connected world, with a continuous increase in the number of IoT devices, the lives of people will become more and more heavily dependent on IoT. In such a hyper-connected world with devices continuously monitoring and assisting the every day lives of people, keeping their lives private is not a very easy task. To do so, IoT systems should be fully compliant with legal requirements and built under the concept of \textit{privacy by default}~\cite{gdpr}. The IoT experimentation world can not be distinguished from the real world, since the data that are gathered by testbeds (that support the experiments) are real-world data belonging to or describing real people. In both cases (real-world implementation and experimentation) a well-defined ontology providing the required privacy-enhancing features to protect the users' data is mandatory. This ontology should describe the entities of an IoT system and their interconnections, embedding also the most important concepts of privacy to ensure that any IoT system is privacy-enhanced. 

Having this in mind, in this work we enhance existing well-defined ontologies (such as that defined in~\cite{Agarwal2016unified}) adding the necessary components to address the main {data capturing and sharing} requirements of GDPR mapped to the specifics of an IoT system. { It has to be noted here that the proposed ontology does not claim to address all the GDPR requirements, especially not the ones related to data processing, retention or deletion, but rather focuses on the privacy requirements for data gathering and sharing}. The ontology addresses the GDPR requirements {described in Section~\ref{sec:req}}, by adding a new subgraph as the \textit{privacy graph}, which includes key concepts as \textit{consent}, \textit{permission}, \textit{data subject}, \textit{purpose} and linking these concepts with IoT services and data controllers. The main goal of this graph is to ensure that any request for data will have to get the permission of the data subject (if the data are \textit{sensitive}). In order to do so, the requester has to provide a well-defined description of the purpose, ensuring that the user will have the full control and full view over who has access to his data. 

It is expected that such an ontology will be highly adopted by IoT experimentation projects, as well as IoT services that first aim to protect the user privacy. Since this is a high-level view of a privacy-enhanced IoT ontology, in the future, we would like to focus on extending our ontology to include,  for example, concepts related to conditions or context in which a particular observation is captured. A context could be where the observations are made such as indoor/outdoor, overground/underground, etc.~\cite{AgarwalMDM2019}. This could be useful for data analytics, outlier detection and Complex Event Processing (CEP) tasks that require knowledge about external conditions.

{We} would like to add even more privacy concepts to address more fine-grained scenarios than might not be fully captured by this version, i.e., specifying in detail how users can themselves change/add new policies/rules, how the system will ensure that user data will not be kept indefinitely (GDPR requires to set a time period for keeping the data) , and how to account for data breaches. Further, we are also working towards standardising our ontology.


\vspace{-.35cm}
\section*{Acknowledgement}
All the authors have contributed equally. Authors would like to thank the members of FIESTA-IoT consortium and the anonymous reviewers for their valuable comments. This work has been sponsored by the EU Horizon 2020 programme through ACTIVAGE (contract no. 732679) and IoTCrawler (contract no. 779852) projects {and by  the  Science  Foundation  Ireland under  the  grant  number SFI/12/RC/2289\_P2.}
\begin{figure}
    \centering
    \includegraphics[width=\columnwidth]{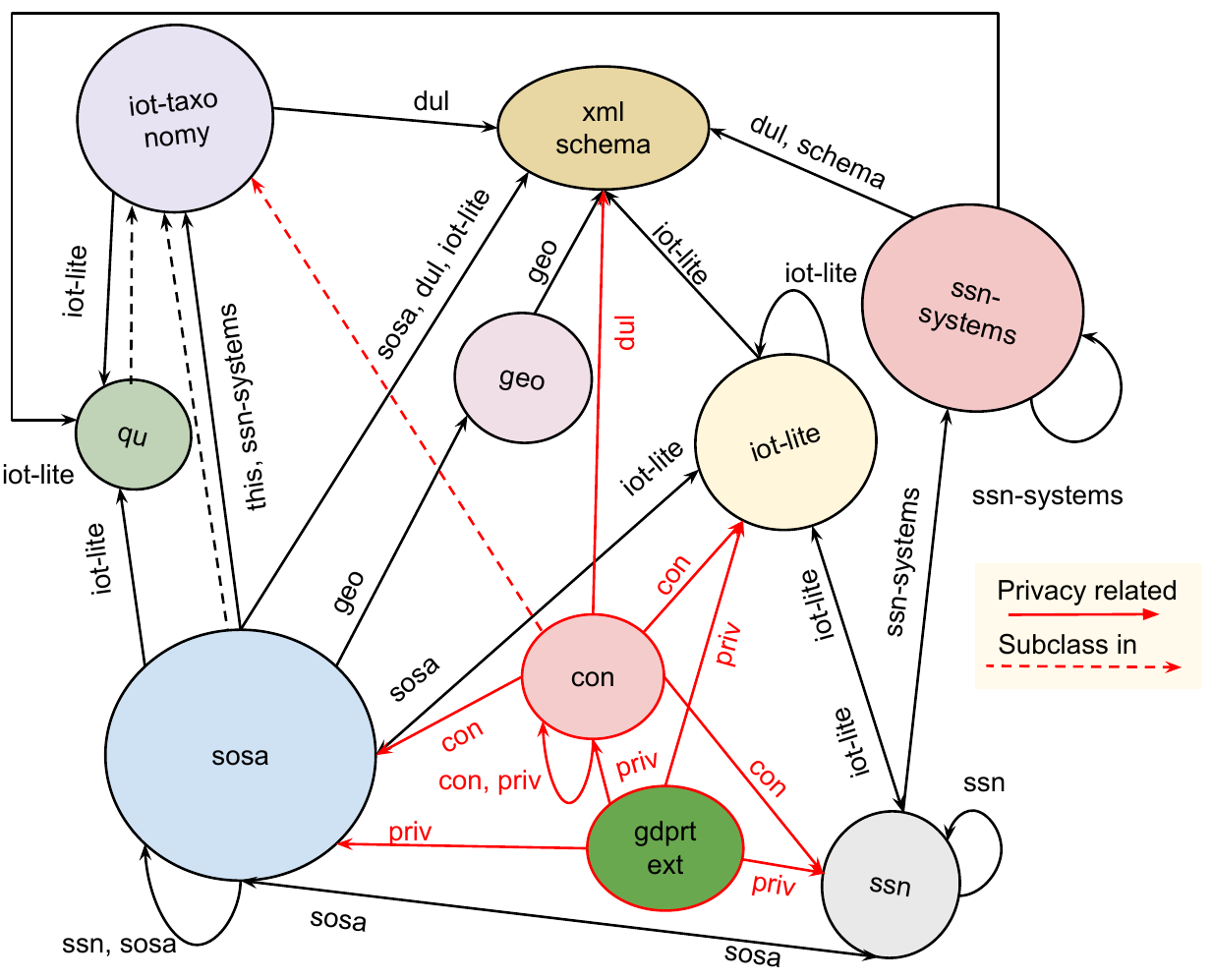}
    \caption{Different Ontologies used. Circles represent different ontologies used. Connections between different ontologies are made using object properties coming from stated ontologies. 
    Privacy-related concepts are shown in red. The direction of the dotted line represents from which ontology the ``subclasses'' for a particular concept is taken from.}
    \label{fig:ontologyConnections}
    \vspace{-0.35cm}
\end{figure}

\begin{figure*}
    \centering
    \includegraphics[width=.82\textwidth]{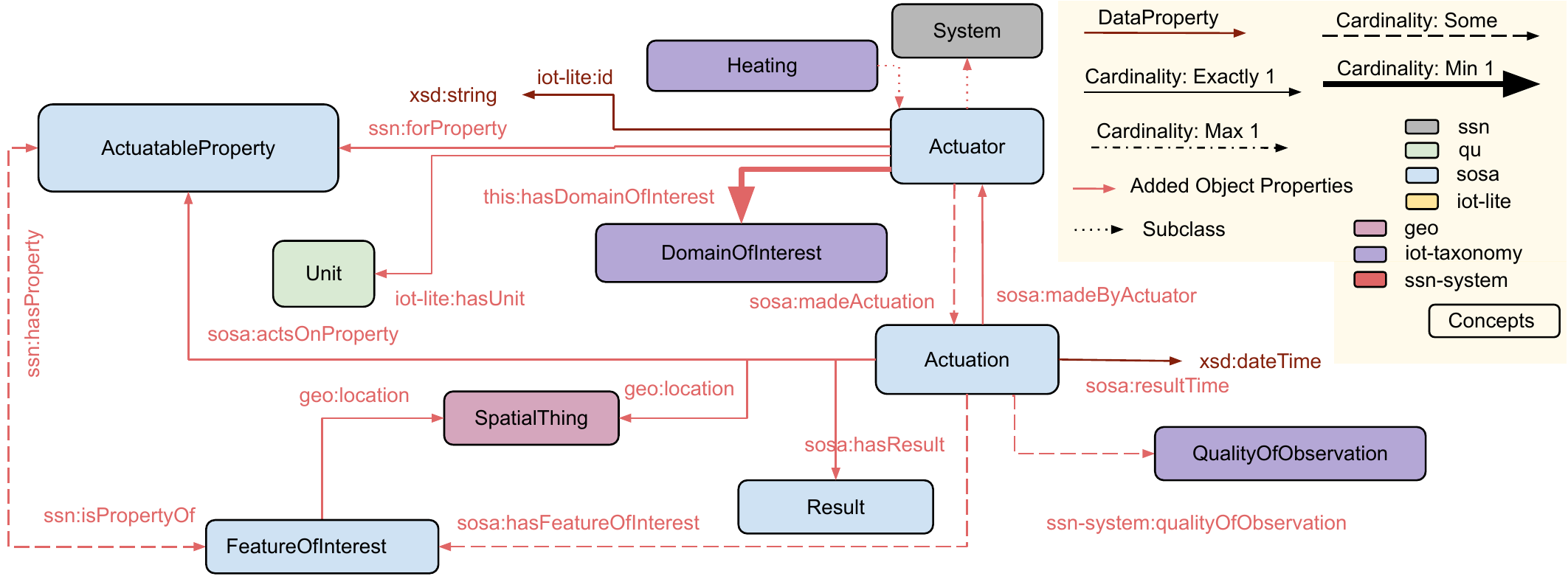}
    \caption{Actuation related ontology graph.}
    \label{fig:actuationGraph}
    \vspace{-0.3cm}
\end{figure*}

\vspace{-0.15cm}
\bibliographystyle{ieeetr}
\bibliography{biblio.bib}

\vspace{-0.25cm}
\appendix
\section{Changes performed}
Figure~\ref{fig:ontologyConnections} shows how we connect different ontologies.
From the actuation point of view, the most relevant concepts and the objectProperties linking them in our ontology are shown in Figure~\ref{fig:actuationGraph}. Note that for the concepts such as \textit{sosa:Results}, its connections are same as in Figure~\ref{fig:ontology}.
\end{document}